\documentclass[aps,longbibliography,showpacs,twocolumn,superscriptaddress,amsmath,amssymb,verbatim]{revtex4-1}

\usepackage[thinlines]{easytable}
\usepackage{graphicx}
\usepackage{lmodern}
\usepackage{epstopdf}
\usepackage[version=4]{mhchem}
\usepackage{amssymb}
\usepackage{dcolumn}
\usepackage{bm}
\usepackage{subfigure}
\usepackage{makecell}
\usepackage{amsmath} 
\usepackage[pdftex,colorlinks=true,citecolor=blue,linkcolor=blue,urlcolor=blue,bookmarks=true]{hyperref}
\usepackage{booktabs}
\usepackage{multirow} 
\usepackage{array} 

\usepackage{color}
\usepackage{textcomp}
\definecolor{red}{rgb}{1,0,0}

\definecolor{blue}{rgb}{0,0,1}

\definecolor{green}{rgb}{0,1,0}

\usepackage{multirow}
\usepackage[utf8]{inputenc}
\begin{document}

\title{Successive magnetic transitions and multiferroicity in layered honeycomb BiCrTeO$_{6}$}
\author{Arkadeb Pal}
\email{pal.arkadeb@gmail.com}
\affiliation{Zernike Institute for Advanced Materials, University of Groningen, 9747 AG Groningen, The Netherlands}
\affiliation{Department of Physics, National Sun Yat-sen University, Kaohsiung 80424, Taiwan}

\author{P. H. Lee}
\affiliation{Department of Physics, National Sun Yat-sen University, Kaohsiung 80424, Taiwan}

\author{J. Khatua}
\affiliation{Department of Physics, Sungkyunkwan University, Suwon 16419, Republic of Korea}

\author{C. W. Wang}
\affiliation{National Synchrotron Radiation Research Center, Hsinchu 300092, Taiwan}

\author{J. Gainza}
\affiliation{European Synchrotron Radiation Facility, 71 Avenue des Martyrs, 38043 Grenoble, France}

\author{A. Fitch}
\affiliation{European Synchrotron Radiation Facility, 71 Avenue des Martyrs, 38043 Grenoble, France}

\author{Thomas J. Hicken}
\affiliation{PSI Center for Neutron and Muon Sciences, 5232 Villigen PSI, Switzerland}

\author{H. Luetkens}
\affiliation{PSI Center for Neutron and Muon Sciences, 5232 Villigen PSI, Switzerland}

\author{Y.-J. Hu}
\affiliation{Institute of Physics, National Yang Ming Chiao Tung University, Hsinchu 300093, Taiwan}

\author{Ajay Tiwari}
\affiliation{Department of Physics, National Sun Yat-sen University, Kaohsiung 80424, Taiwan}

\author{D. Chandrasekhar Kakarla}
\affiliation{Department of Physics, National Sun Yat-sen University, Kaohsiung 80424, Taiwan}

\author{J.-Y. Lin}
\affiliation{Institute of Physics, National Yang Ming Chiao Tung University, Hsinchu 300093, Taiwan}

\author{K. Y. Choi}
\affiliation{Department of Physics, Sungkyunkwan University, Suwon 16419, Republic of Korea}

\author{G. R. Blake}
\email{g.r.blake@rug.nl}
\affiliation{Zernike Institute for Advanced Materials, University of Groningen, 9747 AG Groningen, The Netherlands}

\author{H. D. Yang}
\email{yang@mail.nsysu.edu.tw}
\affiliation{Department of Physics, National Sun Yat-sen University, Kaohsiung 80424, Taiwan}

\date{\today}

	\begin{abstract}
	Low-dimensional magnetic systems based on honeycomb lattices provide a promising platform for exploring exotic quantum phenomena that emerge from the intricate interplay of competing spin, orbital, lattice, and dipolar degrees of freedom. Here, we present a comprehensive study of the layered honeycomb lattice antiferromagnet $\text{BiCrTeO}_{6}$ using magnetization, specific heat, muon spin–relaxation ($\mu$SR) spectroscopy, dielectric, pyrocurrent, and high-resolution synchrotron x-ray diffraction (SXRD) measurements. Our results reveal an array of intriguing and strongly correlated phenomena, which include two successive antiferromagnetic transitions at $T_{\rm N1} \approx 16 \text{ K}$ and $T_{\rm N2} \approx 11 \text{ K}$, a pronounced magnetodielectric coupling effect, and ferroelectric order at  $T_{\rm N2}$. Consequently, this compound emerges as a new spin-driven multiferroic system. The SXRD analysis reveals a magnetoelastic-coupling-induced structural phase transition at $T_{\rm N2}$, characterized by a symmetry lowering from \textit{P}$\overline{3}\,1\,c$ (163) to \textit{P}$3\,1\,c$ (159), which likely triggers the onset of ferroelectricity. In addition to its low-temperature multiferroic behavior, the system exhibits dielectric relaxor characteristics at higher temperatures within the paramagnetic region ($T < 50\,\mathrm{K}$), which is intrinsically linked to the antisite disorder of Cr and Te atoms. 
	\end{abstract}
\maketitle

\section{Introduction}
Magnetic ferroelectrics, also known as multiferroics, are a fascinating class of materials, wherein ferroelectricity appears only in the magnetically ordered state \cite{daniel_2009,annurev,Spaldin2019}. Materials exhibiting multiferroic properties are rare in nature due to the mutually antagonistic origins of magnetic and ferroelectric orders, as well as the stringent symmetry constraints required to sustain both orders. Consequently, the challenge of designing new multiferroic materials with enhanced functionalities persists. 
\\ \hspace*{0.5 CM}  In this regard, exploring low-dimensional spin-frustrated magnetic materials is particularly compelling. The intricate competition among various spin interactions and the inherent frustration, often arising from the unique topology of the spin lattices, can give rise to complex and unconventional spin orders. These intricate spin orders can, in turn, induce ferroelectricity, making such systems especially attractive for multiferroic studies \cite{annurev}. The fascinating interplay among various order parameters, such as spins, charges, and dipoles, serves as a fundamental mechanism driving multiferroicity in various systems. This strong cross-coupling between different ferroic orders opens exciting possibilities for controlling one property through an alternative stimulus, beyond the conventional external fields \cite{Spaldin2019,Eerenstein2006}. This unique property paves the way for advanced spintronic device architectures and diverse real-world applications. It also offers a rich platform for fundamental research, deepening our understanding of the interplay between multiple order parameters \cite{Katsura}.\\
\hspace*{0.5CM}  Geometrically spin-frustrated magnets have garnered significant research interest globally due to their exhibition of a variety of exotic phenomena, which originate from inherent spin frustration and quantum fluctuations. These phenomena include spin liquids, spin ices, and fractionalized excitations such as Majorana quasiparticles and spinons, along with the observation of magnons and multiferroicity \cite{RbFe_white,PhysRevB.90.224402,PhysRevB.86.140405, qsl_science1,CTO_magnons,Sudi_FPO}. In particular, the honeycomb spin system offers a quintessential platform for exploring quantum spin liquid states. This is exemplified by the exactly solvable Kitaev model, which features bond-dependent Ising interactions within a honeycomb lattice. A honeycomb lattice governed by isotropic nearest-neighbor Heisenberg interactions, free from intrinsic geometric frustration, typically hosts a classical Néel-type antiferromagnetic (AFM) long-range order \cite{Honeycomb_theo1}. However, the addition of competing further-neighbor interactions introduces significant frustration into the system, giving rise to a rich variety of magnetic states. These include quantum spin liquids, valence bond states, and spiral spin orderings \cite{valencebond_honeycomb1,Spiral_honeycomb1}. While such honeycomb lattice systems have been extensively explored for their intriguing magnetic properties, research into their multiferroic behavior remains particularly limited, despite their significant potential for hosting multiferroicity.\\ 
\hspace*{0.5CM} In recent years, a few honeycomb lattice magnetic systems have been discovered to exhibit multiferroicity and strong magnetoelectric coupling, driven by their complex and nontrivial spin structures. These findings have sparked a surge of renewed interest and research in this exciting area \cite{Arima_CNO1,Liu_Co2Mo3O8,Shameek_NCTO}. However, the impact of their magnetic topology on ferroelectric order remains poorly understood \cite{RbFe_white}. Thus, further exploration of such systems could lead to the discovery of new and intriguing multiferroics and magnetoelectrics, as well as provide deeper insights into the underlying mechanisms that drive spin-induced ferroelectricity. \\
\hspace*{0.5CM}  In the present report, we delineate a comprehensive study of the physical properties of a layered honeycomb lattice system, $\text{BiCrTeO}_{6}$. This compound was first reported by Vats \textit{et al.} in 2013 \cite{vats2013} and subsequently identified as an antiferromagnet through bulk magnetization measurements by Kim \textit{et al.} \cite{kim2016_BCTO}.  However, the combination of a layered two-dimensional spin-frustrated honeycomb lattice and the presence of \(\mathrm{Bi}^{3+}\) ions with 6s lone pairs makes $\text{BiCrTeO}_{6}$ particularly intriguing for further exploration of its magnetic and multiferroic properties, which have yet to be reported. Here, our current investigations reveal that $\text{BiCrTeO}_{6}$ exhibits multiferroicity, along with a strong magnetodielectric coupling effect triggered by a structural phase transition that closely corresponds to its complex spin ordering.
\begin{figure}[t]
	\centering
	\includegraphics[width = \columnwidth]{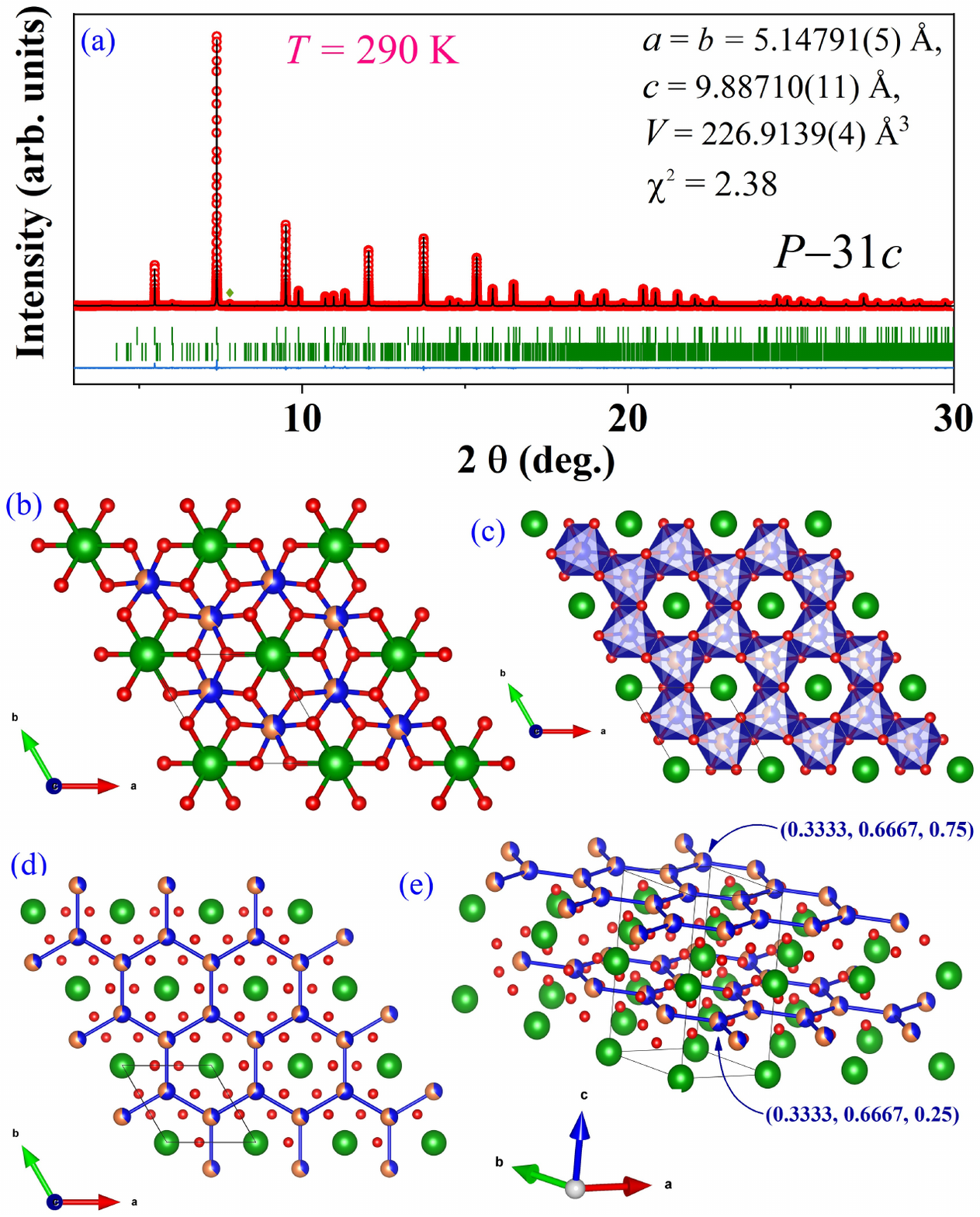}
	\caption{(a) SXRD pattern at $T = \text{290 K}$ (red circles) with the fitted and differences profiles (black and blue lines). Upper and lower green bars indicate the Bragg peak positions for the main phase of BiCrTeO$_6$ and the impurity phase (less than 1\,\text{wt.\%}), Bi$_6$Te$_2$O$_{15}$, respectively. The impurity phase peak is marked with a green $\Diamond$ symbol. (b) Crystal structure of BiCrTeO$_6$ as viewed perpendicular to the \textit{ab} plane. (c) Corresponding polyhedral structure depicting the edge-shared Cr/TeO$_{6}$ octahedra that form the honeycomb structure. (d) Representation of the honeycomb lattice formed by Cr/Te atoms. (e) Two consecutive honeycomb layers stacked along the \textit{c} axis.}
	\label{Fig1}
\end{figure}
\section{EXPERIMENTAL DETAILS}
Polycrystalline BiCrTeO$_6$ powder was prepared using a conventional solid-state synthesis method. High-purity ($>99.99\%$) oxide powders of Bi$_2$O$_3$, Cr$_2$O$_3$, and H$_6$TeO$_6$ were used as precursors, thoroughly mixed in the appropriate stoichiometric ratio, and then ground for 1 h. The mixture was first subjected to heat treatments starting at \(700^\circ\mathrm{C}\) for 24~h under oxygen flow. This was followed by two additional heating steps at 750~$^\circ$C and 780~$^\circ$C, each lasting 12 h under oxygen flow, with intermediate grinding and pressing into pellet forms to ensure homogeneity. To reduce antisite disorder, the sample was slowly cooled to ambient temperature.\\
\hspace*{0.5CM}The phase purity and detailed crystal structure of the synthesized material were examined using high-resolution synchrotron X-ray diffraction (SXRD) at the ID22 beamline at the European Synchrotron Radiation Facility (ESRF) in Grenoble, France \cite{Fitch:vy5010}. The X-ray wavelength employed was 0.42622~\AA. The sample was first cooled to the base temperature of 4~K using a helium-based cryostat. Subsequently, the SXRD data were recorded during heating the sample, with approximately 15 minutes of waiting time at each temperature to ensure thermal stability before data collection. Data were collected in a temperature range of 4~K to 290~K. The SXRD data were refined using the FullProf Suite software package \cite{FPsuite}.\\ \hspace*{0.5CM}Magnetic measurements were carried out using a commercial SQUID magnetometer (MPMS, Quantum Design). \(T\)-dependent specific-heat (\(C_{p}\)) measurements were performed using a heat-pulsed thermal relaxation calorimeter integrated into a Physical Property Measurement System (PPMS).\\ \hspace*{0.5CM}The dielectric properties were measured with an Agilent E4980A LCR meter connected to the PPMS system using a custom dielectric probe. Pyrocurrent measurements were performed with a Keithley 6517A electrometer, while temperature and magnetic field parameters were controlled via the PPMS system.\\ \hspace*{0.5CM} Zero-field (ZF) and weak-transverse-field (wTF)  muon spin relaxation ($\mu$SR) experiments were conducted down to 1.5 K on the GPS instrument at the Swiss Muon Source, Paul Scherrer Institute in Villigen, Switzerland \cite{GPS_2}. Approximately 0.5 grams of the polycrystalline powder sample was placed into a silver foil packet measuring about \(1.2 \times 1.2 \, \text{cm}^2\) and secured to the sample fork. This assembly was then inserted into the muon beam chamber, where a fully spin-polarized muon beam was directed at the sample. The system was cooled to a base temperature of 1.5 K using a liquid helium cryostat. Data collection was conducted at various temperatures while the sample was heated. The collected $\mu$SR data were analyzed using the MUSRFIT software \cite{MUSRFIT}.

\section{RESULTS AND DISCUSSION}

\subsection{Crystal structure}
Figure \ref{Fig1} (a) presents the experimental SXRD pattern collected at $T = \text{290 K}$ alongside the fitted and difference profiles obtained by Rietveld refinement. The data were successfully fitted assuming a trigonal structure with space group \(P\bar{3}1c\). The crystallographic parameters obtained are summarized in Table SI in the supplemental materials (SM) \cite{Supplemental}, which is in good agreement with previous reports \cite{kim2016_BCTO,vats2013}.
A trace of a tiny impurity phase of Bi$_6$Te$_2$O$_{15}$, marked with a green $\Diamond$ symbol and constituting less than 1\,\text{wt.\%}, was detected in the SXRD data. This impurity phase was unavoidable during sample preparation, and similar unidentified impurity phases have also been reported in earlier studies \cite{kim2016_BCTO}. In a fully ordered structure, Cr and Te atoms are expected to occupy distinct sites characterized by the 2\textit{d} (0.6667, 0.3333, 0.25) and 2\textit{c} (0.3333, 0.6667, 0.25) Wyckoff positions, respectively. A high degree of cationic ordering would produce a prominent (1 0 1) Bragg peak around $2\theta = 6^\circ$. However, in the observed SXRD pattern [Fig. \ref{Fig1} (a)], the (1 0 1) peak is very weak. This strongly indicates the presence of significant antisite disorder, arising from the random exchange of Cr and Te atoms between their respective lattice sites in BiCrTeO$_6$. Further analysis of the SXRD data revealed a site occupancy of Cr : Te = 0.62 : 0.38 at the 2\textit{d} site, and Te : Cr = 0.62 : 0.38 at the 2\textit{c} site. This substantial antisite disorder can be attributed to the similar ionic radii of \(\mathrm{Cr}^{3+}\) [$r_{io} = 0.615\,\mathrm{\AA}$] and \(\mathrm{Te}^{6+}\) [$r_{io} = 0.56\,\mathrm{\AA}$], which facilitates site exchange between these ions.

The crystal structure of BiCrTeO$_6$ is schematically depicted in Figs. \ref{Fig1} (b)–(e). It consists of two-dimensional (2D) honeycomb layers in the \(ab\)-plane, comprised of edge-shared CrO\(_6\) and TeO\(_6\) distorted octahedra, as shown in Figs. \ref{Fig1}(b) and (c). The resulting honeycomb lattice formed by the Cr/Te atoms is schematically shown in Fig. \ref{Fig1} (d). Two successive honeycomb layers are separated along the crystallographic \(c\)-axis by a distance of \(0.5c\) [Fig.\ref{Fig1}(e)].  Between these Cr/Te honeycomb layers, a 2D layer of Bi atoms is sandwiched. In general, antisite disorder can profoundly influence the physical properties of a material \cite{Blasco_YCMO, PCFO_APL}. Therefore, the site-disordered, spin-frustrated honeycomb lattice in BiCrTeO$_6$ may give rise to more complex magnetic exchange interactions within the system.


\begin{figure}[h!]
\centering
\includegraphics[width = \columnwidth]{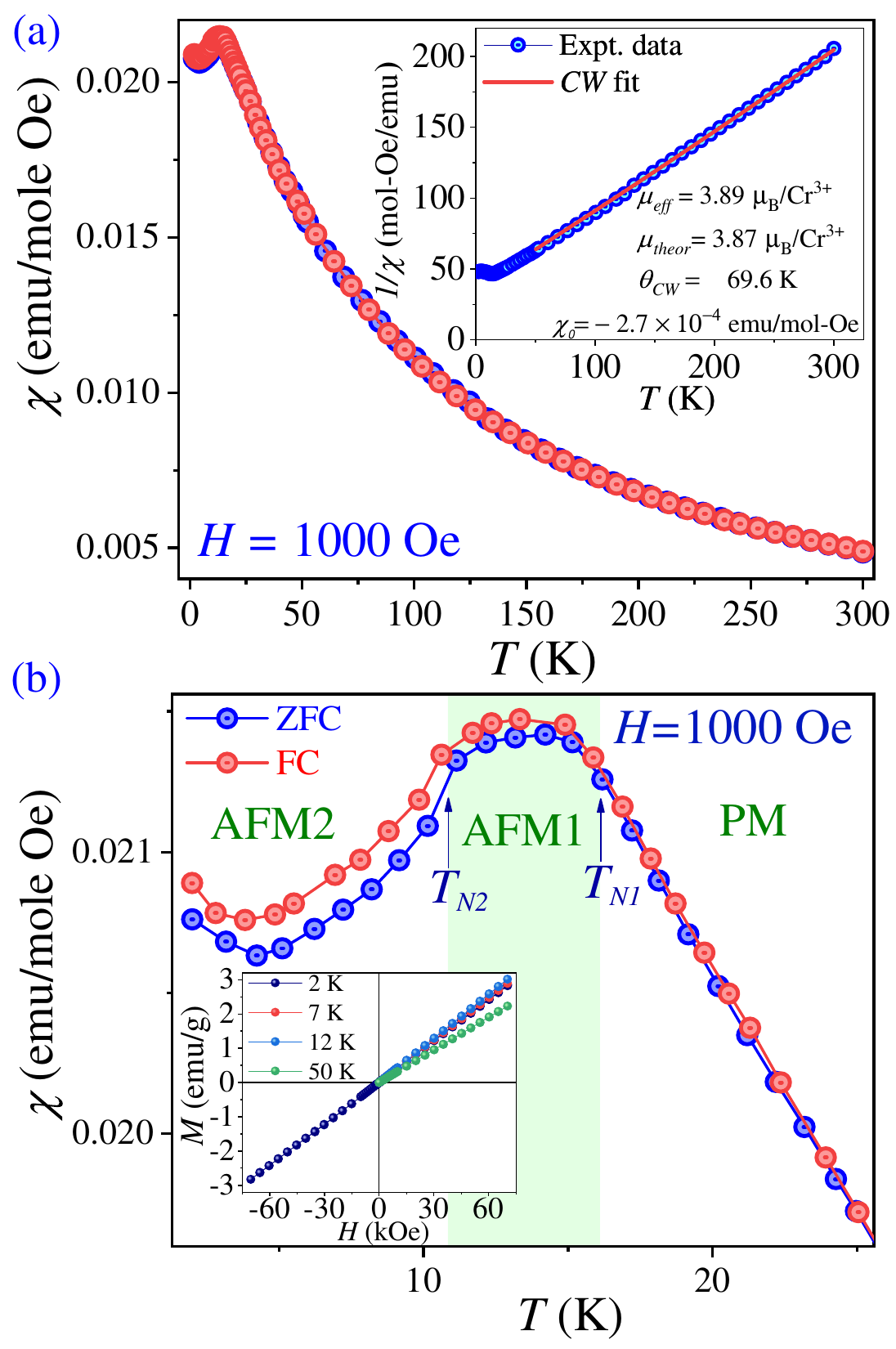}
\caption{(a) and (b) depicts $\chi$($T$) ZFC (shown in blue) and FC (shown in red) curves under \(H\)= 1000 Oe and its closer view near transition temperatures, respectively.  The corresponding Curie-Weiss fitting of “$1/\chi$ vs. $T$” curve is shown in the inset of Fig. (a). \(M\)(\(H\)) curves collected at \(T\) = 2, 7, 12, and 50 K are shown in the inset of Fig.(b).}
\label{Fig2}
\end{figure}

\subsection{Magnetic properties}\label{CHI}
Figure \ref{Fig2} (a) displays the temperature dependence of the magnetic susceptibility ($\chi$) of BiCrTeO$_6$ under an applied field of \(H= 1000\) Oe following the standard protocols of field cooling (FC) and zero-field cooling (ZFC).
The $\chi$($T$) curve exhibits a broad peak, defined by changes in slope at $T_{\rm N1} \approx 16 \text{ K}$ and $T_{\rm N2} \approx 11 \text{ K}$. The broad nature of the peak is more evident in a closer view of the $\chi$($T$) curve in this temperature region, as displayed in Fig. \ref{Fig2} (b). In a previous study on BiCrTeO$_6$, a broad peak observed in the $\chi$($T$) curves near 17 K was reported, which was taken as evidence of an antiferromagnetic (AFM) spin ordering \cite{kim2016_BCTO}. However, the broadness of this peak makes it difficult to unambiguously determine the exact transition temperature. Therefore, the nature of the magnetic ordering requires additional confirmation through independent and more sensitive experimental probes. In fact, our investigation using \(C_{p}\) and $\mu$SR measurements clearly demonstrates the occurrence of two successive transitions at  $T_{\rm N1} \approx 16 \text{ K}$ and $T_{\rm N2} \approx 11 \text{ K}$, which will be discussed in the following sections. Notably, other similar compounds, such as RCrTeO$_6$ (R=Y, La, Tb, Er) and BiMnTeO$_6$, have been reported to exhibit AFM ordering at temperatures near \(T_{\rm N2} \approx 11\) K \cite{kim2016_BCTO,RCTO1}. 

The high-temperature paramagnetic susceptibility ($\chi$) was fitted using the modified Curie - Weiss (CW) law:
\begin{equation}
\chi = \chi_0 + \frac{C}{T - \theta_{\mathrm{CW}}}
\end{equation}
where $\chi_0$ comprises the core diamagnetism associated with the core-electron shells ($\chi_{\mathrm{core}}$) of the constituent ions and the van Vleck paramagnetism ($\chi_{\mathrm{VV}}$) related to the open shells; $C$ is the Curie constant, and $\theta_{\mathrm{CW}}$ is the CW temperature.  
The inset of Fig.~\ref{Fig2}(a) depicts the best fit, which yielded $\chi_0 = -2.7 \times 10^{-4}$ emu/mol-Oe, $\mu_\mathrm{eff} = 3.89\, \mu_B/$Cr$^{3+}$, and $\theta_{\mathrm{CW}} = -69.6$~K.  
The negative $\theta_{\mathrm{CW}}$ indicates the presence of dominant AFM interactions.  
The estimated $\mu_\mathrm{eff} = 3.89\, \mu_B/$Cr$^{3+}$ aligns well with the calculated spin-only moment of Cr$^{3+}$ ions (~3.87 $\mu_B$).

The isothermal magnetization (\textit{M}) as a function of applied magnetic field (\textit{H}) at various temperatures is shown in the inset of Fig.~\ref{Fig2}(b). The linear and non-saturating behavior of the \(M\)(\(H\)) curves below $T_{\rm N1}$ and $T_{\rm N2}$ indicates the AFM nature of the magnetic ordering.

\subsection{Specific heat}\label{HC}
\begin{figure}[h!]
\centering
\includegraphics[width = \columnwidth]{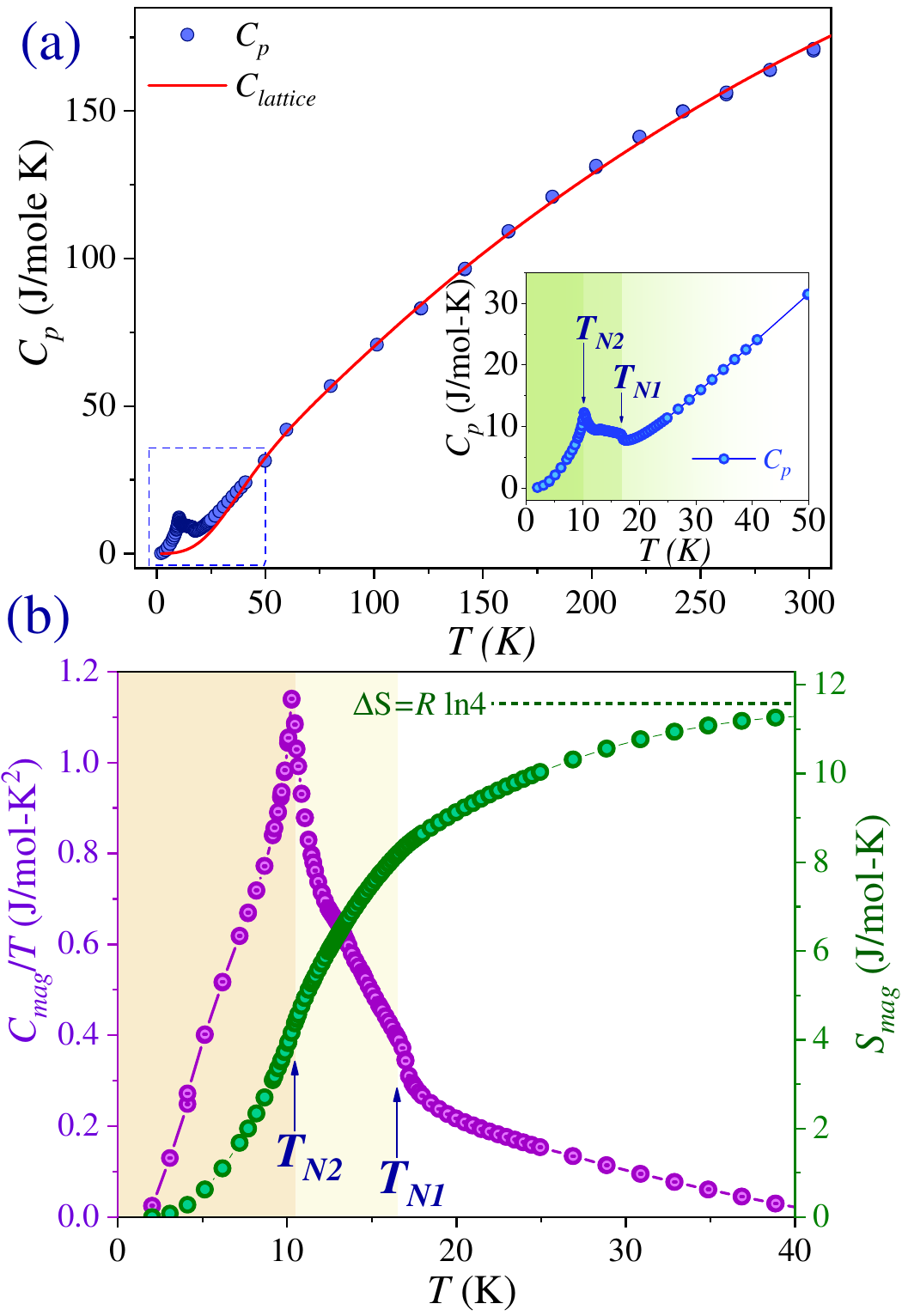}
\caption{(a) Temperature dependence of the total specific heat \(C_p(T)\) (blue circles) along with the fitted phononic contribution  using the Debye–Einstein model (red solid line). The inset shows an expanded view at low temperature. (b) \(T\)-variation of \(C_{\text{mag}}\)/$T$ (right $y$-axis) and the corresponding magnetic entropy change (\(S_{\text{mag}}\); left $y$-axis). Long-range magnetic ordering is demonstrated by the \(\lambda\)-like anomaly at \(T_{N2} \approx 11\) K, whereas the shoulder-like anomaly near \(T_{\rm N1} \approx 16\) K indicates another magnetic transition.}
\label{Fig3}
\end{figure}

Figure \ref{Fig3}(a) shows the specific heat (\(C_p\)) versus temperature of BiCrTeO$_6$ collected under \(H=0\) Oe, which displays a distinct \(\lambda\)-like feature at \(T_{\rm N2} \approx 11\) K, thus suggesting the onset of long-range magnetic ordering. In addition, a shoulder-like feature near \(T_{\rm N1} \approx 16\) K is observed, as evident from an expanded view of \(C_p(T)\) in the inset of Fig. \ref{Fig3}(a), suggesting the presence of a second magnetic transition. It should be noted that our $\mu$SR measurements also reveal two successive magnetic transitions (to be discussed later) at \(T_{\rm N1} \approx 16\) K and \(T_{\rm N2} \approx 11\) K, thus corroborating the specific heat and magnetization results. In magnetic insulators, \(C_p(T)\) emanates from the phononic contribution in the high-temperature region, whereas the main contribution is of magnetic origin in the low-temperature region. To extract the magnetic specific heat (\(C_{\text{mag}}\)), we subtracted the phononic contribution (\(C_{\text{lattice}}\)) from the total \(C_p\).

 The lattice contribution \(C_{\text{lattice}}\) was approximated from the fitting of our \(C_p(T)\) data in the elevated temperature regime (30 - 300 K), which was subsequently extrapolated down to 2 K, as shown by the red line in Fig. \ref{Fig3}(a). The fit employed the Debye-Einstein model, expressed as follows:\\
$C_{\text{lattice}}(T) = n_D \left[ 9 R \left(\frac{T}{\theta_D}\right)^3 \int_0^{\theta_D / T} \frac{x^4 e^{x}}{(e^{x} - 1)^2} dx \right]
+ \sum_{i=1}^{3} 3 n_{Ei} R \left(\frac{\theta_{E_i}}{T}\right)^2 \frac{e^{\theta_{E_i}/T}}{\left(e^{\theta_{E_i}/T} - 1 \right)^2}$;
\\ \hspace*{0.5 CM} Here, the first and second terms refer to the acoustic phonons (Debye part) and optical phonons (Einstein part), respectively, wherein \( n_D \) and \( n_{Ei} \) represent their respective weighting factors. \( \theta_D \) and \( \theta_{Ei} \) denote the Debye and Einstein temperatures, respectively. During the fit, the coefficient of the Debye term was initially fixed to \( n_{D} = 3 \), corresponding to the three acoustic modes. The total Einstein weight \( \sum_i n_{E_i} \) was then distributed among the optical branches so that it matched the expected \( 3n - 3 \) optical modes, where n is the total number of atoms in BiCrTeO$_{6}$ formula unit. The best overall fit was obtained using a  Debye branch with \( n_{D} = 3 \) and \( \Theta_{D} = 224~\text{K} \), together with three Einstein branches: \( n_{E1} = 3 \) with \( \Theta_{E1} = 552~\text{K} \), \( n_{E2} =8 \) with \( \Theta_{E2} = 580~\text{K} \), and \( n_{E3} = 13 \) with \( \Theta_{E3} = 1230~\text{K} \). 
\\ \hspace*{0.5 CM} After subtracting the lattice contribution and dividing the resulting magnetic specific heat ($C_{mag}$) by temperature, two anomalies are distinctly revealed at \(T_{\rm N1} \approx 16\) K and \(T_{\rm N2} \approx 11\) K, as shown on the left axis of Fig. \ref{Fig3}(b). The obtained \( C_{\text{mag}}/T \) curve as a function of \( T \) was integrated to estimate the magnetic entropy \( S_{\text{mag}} \), as shown in Fig. \ref{Fig3}(b) (right $y$-axis). The estimated saturation value of \( S_{\text{mag}} \) is found to be \(\sim 11.3\,\text{J/mol-K}\), which is close to the total expected magnetic entropy \(\sim 11.5\,\text{J/mol-K}\) for \(\mathrm{Cr}^{3+}\) spins (\(S=3/2\)).

\begin{figure*}[htbp]
    \centering
    \includegraphics[width=\textwidth]{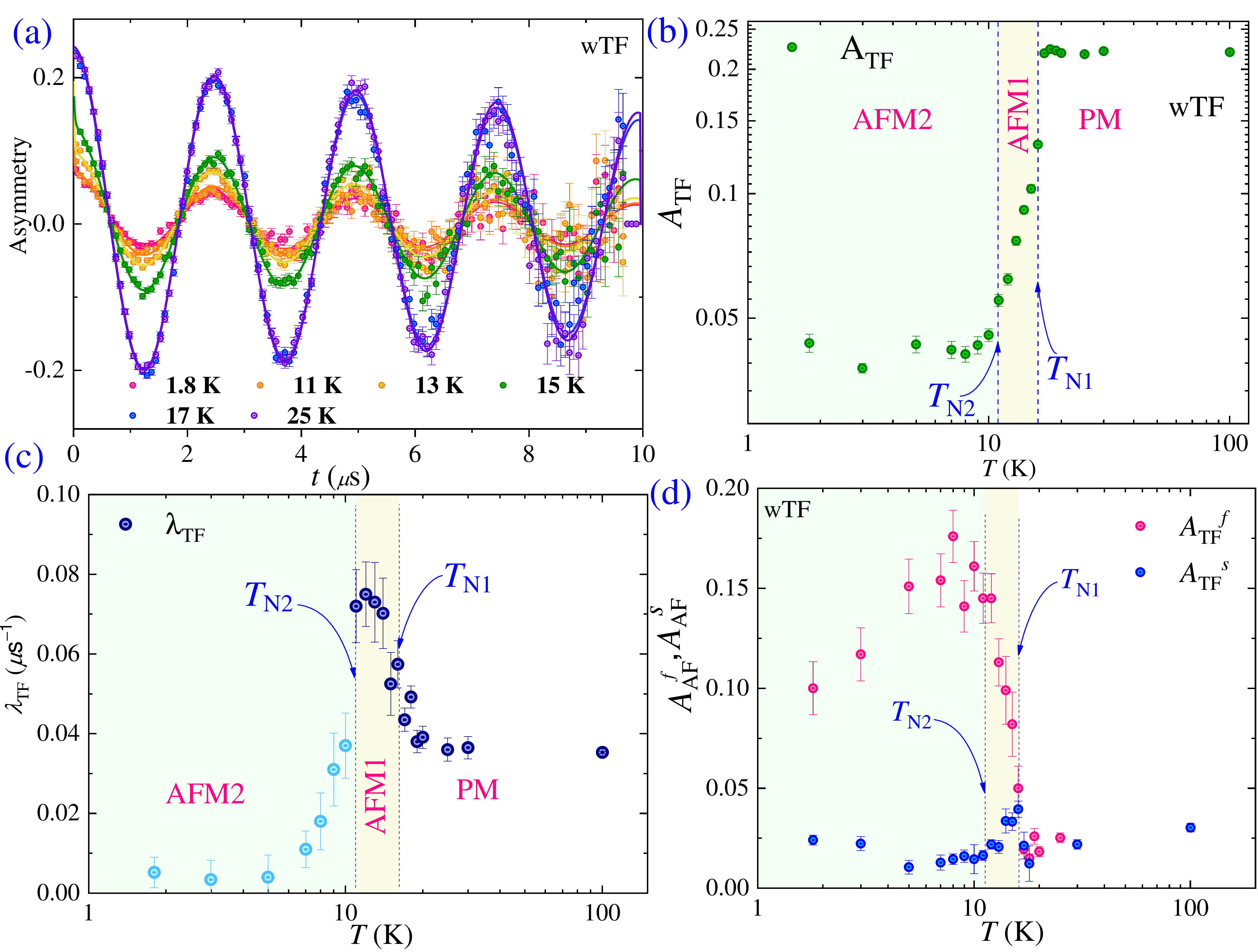}
    \caption{
(a) Time evolution of the muon spin asymmetry collected at various selected temperatures under an applied weak transverse field $H_{\rm wTF}=30$~Oe. The solid lines refer to the theoretical fits as described in the text. Panel (b) shows the temperature dependence of the wTF-asymmetry $A_{\rm TF}$, while panel (c) presents the variation of $\lambda_{\rm TF}$ with temperature. (d) depicts thermal variations of $A_{\mathrm{TF}}^{f}$ and $A_{\mathrm{TF}}^{s}$. Dashed vertical lines mark the transition temperatures $T_{\rm N1}$ and $T_{\rm N2}$.
}
    \label{fig4}
\end{figure*}

\subsection{Muon spin relaxation study}
\begin{figure*}[htbp]
    \centering
    \includegraphics[width=\textwidth]{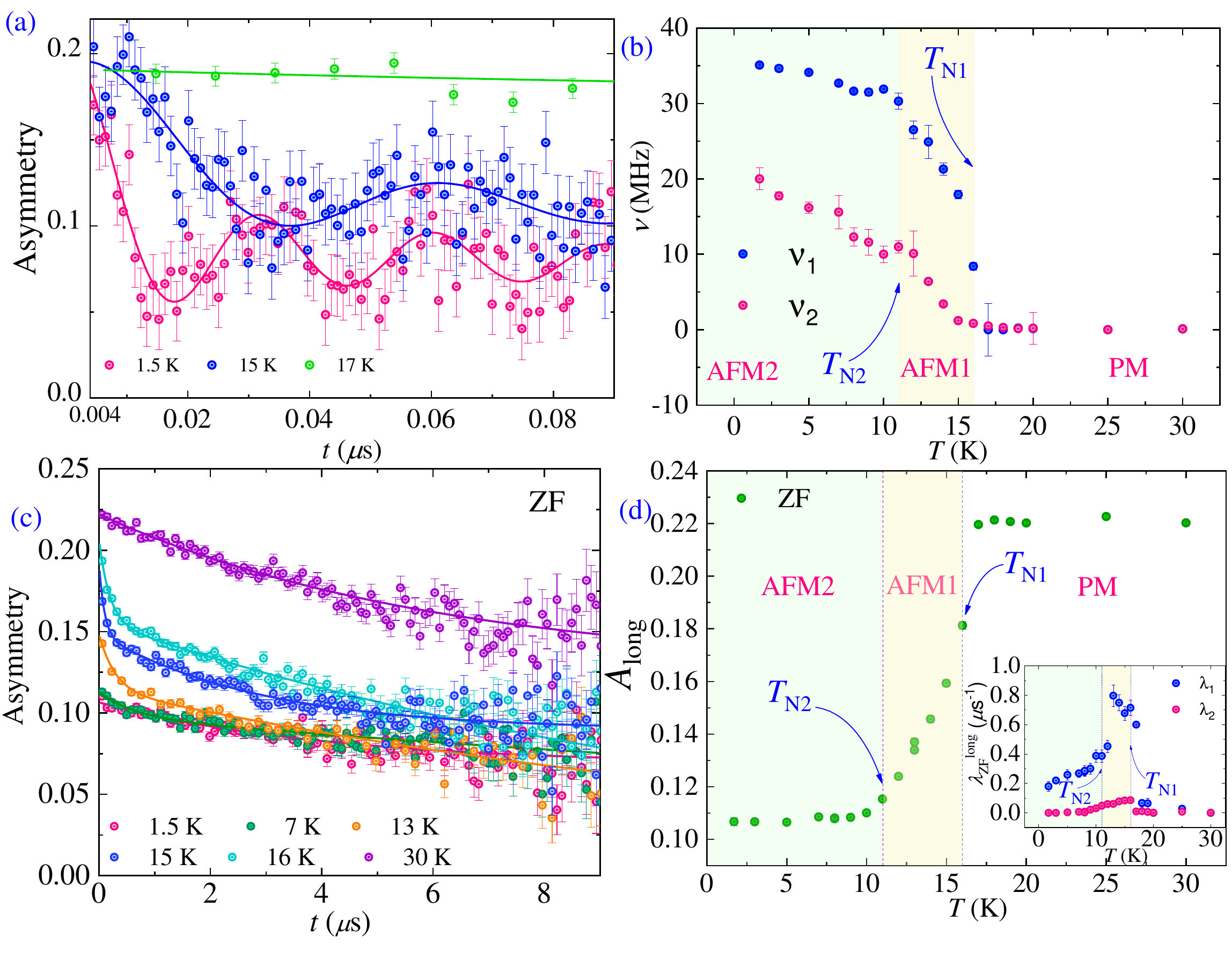}
    \caption{(a) Zero-field (ZF) muon spin asymmetry of \(\mathrm{BiCrTeO_6}\) at shorter time scales for selected temperatures above and below $T_{\rm N1}$ and $T_{\rm N2}$. The solid lines represent the corresponding fits using the model described in the text. (b) Temperature dependence of the two oscillation frequencies, which likely correspond to different internal fields sensed by muons below $T_{\rm N1}$. (c) Time dependence of the ZF muon spin asymmetry at longer time scales for several representative temperatures. The solid lines correspond to the sum of exponential functions as described in the text. (d) Temperature dependence of the asymmetry associated with the non-oscillatory component observed at longer time scales. Inset: Temperature dependence of the fast ($\lambda_{\mathrm{ZF}}^{f}$) and slow ($\lambda_{\mathrm{ZF}}^{s}$) relaxation rates at longer time scale.}
    \label{Fig5}
\end{figure*}
The high gyromagnetic ratio of muons, combined with the fully spin-polarized beams, makes $\mu$SR a uniquely powerful technique for probing static and dynamic spin properties at the microscopic level~\cite{Hillier2022}. Hence, we performed ZF and wTF-$\mu$SR measurements on polycrystalline sample of \(\mathrm{BiCrTeO_6}\) to gain deeper insight into its magnetic properties, providing a complementary perspective to the bulk characterization discussed in Sects.~\ref{CHI} and \ref{HC}.\\ \hspace*{0.5 CM} Figure~\ref{fig4}(a) presents the time evolution of the muon-spin asymmetry at several temperatures under a wTF of $H_{\rm wTF}=30$~Oe, applied perpendicular to the initial muon-spin direction.
 The oscillatory feature of the asymmetry observed at temperatures $T>$ $T_{\mathrm{N1}}$ indicates that the muon spins precess coherently around the applied transverse field with a frequency $\gamma_{\mu} \mu_0 H_{\mathrm{wTF}} / 2\pi$, where $\gamma_{\mu} = 2\pi \times 135.5~\mathrm{MHz/T}$ is the muon gyromagnetic ratio. This behavior is typical of paramagnetic states with fast fluctuating electronic spins and randomly oriented nuclear spins that are static on the muon timescale \cite{muon_basic1,Blundell_muon}. \\ \hspace*{0.5 CM}As the sample develops long-range magnetic order below $T_{\mathrm{N1}}$, the strong internal magnetic fields begin to dominate the muon spin precession, effectively exceeding the weak applied external field. Consequently, the oscillation amplitude is progressively suppressed and usually disappears once the system becomes fully magnetically ordered. Upon cooling below \(T_{\mathrm{N1}}\), the asymmetry exhibits a reduced oscillation amplitude that persists over longer time scales, while a fast relaxation component appears at shorter times. Down to $T$ = 2 K,  the weak yet discernible oscillatory background signal may suggest the presence of persistent spin dynamics or muons that are stopping in the sample holder \cite{muon_partial2,muon_partial3}. The initial rapid loss of asymmetry in the wTF-$\mu$SR time spectra observed below \(T_{\mathrm{N1}}\) can be attributed to the development of strong static internal fields resulting from spin ordering within the sample \cite{muon_partial1,muon_partial2,muon_partial3}. However, a loss of asymmetry in the wTF time spectra may also arise from strong relaxation of the dynamically fluctuating spins in the system. Consequently, an unambiguous confirmation of the formation of the magnetic ordering in \(\mathrm{BiCrTeO_6}\) was obtained from further  ZF-$\mu$SR measurements (discussed in a following section).

\hspace*{0.5 CM}To track the transition temperature, the  wTF-$\mu$SR spectra were analyzed using an oscillatory component along with two non-oscillatory depolarizing components, which account for static local moments aligned parallel to the initial muon polarization. The fitting function can be expressed as:
$A_{\rm wTF}(t) = A_{\mathrm{TF}} \cos(2\pi\nu t + \phi)\,e^{-\lambda_{\mathrm{TF}} t} 
+ A_{\mathrm{TF}}^{f}\,e^{-\lambda_{\mathrm{TF}}^{f} t} 
+ A_{\mathrm{TF}}^{s}\,e^{-\lambda_{\mathrm{TF}}^{s} t}$,
where, in the first term, \(A_{\mathrm{TF}}\) represents the asymmetry of the oscillating signal originating from the paramagnetic fraction, \(\phi\) is the phase factor, \(\nu\) is the muon precession frequency,
 and \(\lambda_{\mathrm{TF}}\) denotes the muon spin relaxation rate in \(H_{\mathrm{wTF}}\) = 30 Oe. 
The second and third terms correspond to the fast (pertaining to the fast damping in the early segment of the spectra) and slow relaxation components arising due to any possible relaxations within the system, characterized by the relaxation rates \(\lambda_{\mathrm{TF}}^{f}\) 
and \(\lambda_{\mathrm{TF}}^{s}\), and their respective asymmetries \(A_{\mathrm{TF}}^{f}\) and \(A_{\mathrm{TF}}^{s}\).

A similar wTF-\(\mu\)SR spectrum has also been observed in several frustrated spin systems \cite{muon_partial1,muon_partial2, muon_partial3}. Similarly to the wTF data, three exponential components are also identified in the ZF spectra (as discussed in a following section), completing our wTF fitting model. The background contribution identified in the zero-field data also precesses in the applied field and gives rise to a weakly relaxing oscillatory signal. We therefore absorb this background contribution into the cosine term of the wTF fitting function in order to minimize the number of fitting parameters.\\ \hspace*{0.5 CM} The solid lines in Fig.~\ref{fig4}(a) represent the fit to the wTF-$\mu$SR asymmetry spectra for BiCrTeO$_{6}$. The temperature dependence of the fit parameters $A_{\mathrm{TF}}$ and $\lambda_{\mathrm{TF}}$, obtained for the oscillatory component, are demonstrated in Fig.~\ref{fig4} (b) and Fig.~\ref{fig4}(c). Above $T_{\mathrm{N1}}$, the transverse-field asymmetry $A_{\mathrm{TF}}$ remains nearly temperature-independent with $A_{\rm TF}$ $\approx 0.22$, indicating that the fraction of paramagnetic spins dominates the observed wTF-$\mu$SR signal.\\
\hspace*{0.5 CM} Upon cooling below \(T_{\mathrm{N1}}\), \(A_{\mathrm{TF}}\) begins to decrease, indicating the gradual development of spin correlations and the onset of magnetic ordering. Accordingly, the combined amplitude of the non-oscillatory components $A_{\mathrm{TF}}^{f} + A_{\mathrm{TF}}^{s}$ increases [Fig.~\ref{fig4}(d)] as the system enters the ordered state. Notably, the \(A_{\mathrm{TF}}(T)\) curve exhibits a monotonic decrease as the temperature is lowered from \(T_{\mathrm{N1}} = 16~\mathrm{K}\) down to \(T_{\mathrm{N2}} = 11~\mathrm{K}\). 
Below \(T_{\mathrm{N2}}\), \(A_{\mathrm{TF}}\) reaches a nearly temperature-independent value, implying the presence of a more static ordered state. Moreover, the wTF-$\lambda_{\mathrm{TF}}$ exhibits a pronounced increase as the temperature drops below $T_{\rm N1} = 16$ K. This muon spin-relaxation rate reaches a maximum near $T_{\rm N2} = 11$ K. Such a trend is indicative of a broadening in the internal field distribution and a slowing down of magnetic spin fluctuations close to the critical temperatures. As the applied external magnetic field is weak relative to the internal fields, $\lambda_{\mathrm{TF}}$ values below $T_{\rm N2}$ are generally not considered. Nevertheless, a small residual relaxation remains down to the lowest temperature, which we attribute to weak relaxation of the background signal and/or slowly fluctuating spins persisting within the magnetically ordered state. \\
\hspace*{0.5 CM} To further investigate the nature of static spin ordering, the thermal evolution of the local magnetic fields, and the potential presence of spin dynamics, we now discuss the ZF-$\mu$SR results. In long-range ordered magnetic systems below the transition temperature, muon spins primarily sense the internal static magnetic fields, leading to a characteristic precession signal that manifests as oscillations at short time scales~\cite{KhatuaBMTO,Hillier2022}. The ZF-$\mu$SR spectra of BiCrTeO$_6$, as shown in Fig.~\ref{Fig5}(a), display clear coherent oscillations at short time scales ($t < 0.1~\mu\text{s}$) for $T < T_{\mathrm{N1}}$, confirming the presence of long-range magnetic order in $\mathrm{BiCrTeO_6}$.\\
 \hspace*{0.5 CM} Initial attempts to fit the ZF data, especially in the early-time region ($t < 0.1~\mu$s), using a sum of simple exponentially damped cosine functions proved to be unsatisfactory. After exploring various other oscillatory functions, a satisfactory fit to the data, as depicted by solid lines in Fig.~\ref{Fig5} (a), was obtained by employing a combination of two spherical Bessel functions, $j_{0}$, together with additional exponential components, which is expressed as~\cite{le2011muon} $
  A_{\rm ZF}^{\rm short} (t) = A_{\text{B1}} \, j_0(2 \pi \nu_1 t) \, e^{-\lambda_{\text{B1}} t} + A_{\text{B2}} \, j_0(2 \pi \nu_2 t) \, e^{-\lambda_{\text{B2}} t} + A_{\text{tail}} \, e^{-\lambda_{\text{tail}} t} + A_{\text{BG}} \, e^{-\lambda_{\text{BG}} t}.$  
Here, $A_{\mathrm{B1}}$ and $A_{\mathrm{B2}}$ correspond to the asymmetries associated with the oscillatory components characterized by the frequencies $\nu_{1}$ and $\nu_{2}$. The term $A_{\mathrm{tail}}$ denotes the asymmetry of the non-oscillatory tail produced by muon spins interacting with the components of the local internal magnetic field that are parallel to the initial direction of muon spin polarization. The term $A_{\mathrm{BG}}$ represents a small background contribution, likely originating from muons stopping in the sample holder.
 Moreover, $\lambda_{\mathrm{B1}}$, $\lambda_{\mathrm{B2}}$, $\lambda_{\mathrm{tail}}$, and $\lambda_{\mathrm{BG}}$ correspond to the depolarization rates associated with their respective polarization components. The fit to the experimental data at short time scales, obtained using this model, is shown by the solid line in Fig.~\ref{Fig5}(a).\\ 
 \hspace*{0.5 CM} The fit obtained using this combination of two Bessel functions reveals two oscillatory components, suggesting the possible presence of either two magnetically inequivalent muon stopping environments or magnetic inequivalence arising from symmetry breaking induced by magnetic anisotropy \cite{Two_muon_sites2,Two_muon_sites1,mn5mtkm5,doi:10.7566/JPSJ.89.064703}. This further points to a complex local magnetic environment and potentially a complex magnetic ground state. Before proceeding, we comment that Bessel-function–like muon polarization is frequently encountered in materials hosting complex or noncollinear magnetic structures, such as helical, spiral, or incommensurate antiferromagnets, where the internal magnetic fields experienced by the muons form a broad or continuously distributed spectrum rather than a single unique local field value~\cite{Two_muon_sites2,PhysRevB.89.184425,PhysRevB.106.144401}.\\
 \hspace*{0.5 CM} The thermal variation of the oscillation frequencies $\nu_{\mathrm{1}}$ and $\nu_{\mathrm{2}}$, which are directly proportional to the static internal local magnetic fields, is depicted in Fig.~\ref{Fig5} (b).
 The temperature dependence of both the parameters $\nu_{\mathrm{1}}$ and $\nu_{\mathrm{2}}$ exhibits a similar trend, as they both show a gradual increase below $T_{\rm N1} = 16$ K, indicating the onset of quasi-static magnetic ordering. When the temperature is further lowered below $T_{\rm N2} = 11$ K, an additional kink appears, suggesting another magnetic transition occurring at this temperature.\\ \hspace*{0.5 CM} For longer time scales (\( t > 0.1~\mu\mathrm{s} \)), as shown in Fig.~\ref{Fig5}(c), the rapid oscillations related to static internal fields are no longer discernible because they are effectively averaged out in this visualization due to the coarser time resolution. Our analysis shows that the long-time ZF asymmetry is well described by a two-component exponential model, as illustrated by the solid lines in Fig.~\ref{Fig5}(c). The model can be expressed as follows: $
A_{\rm ZF}^{\rm long}(t)
= A_{\text{long}} [f\, e^{-\lambda_{\rm ZF}^{f} t} + (1 - f)\, e^{-\lambda_{\rm ZF}^{s} t}],
 $
where $\lambda_{\rm ZF}^{f}$ and $\lambda_{\rm ZF}^{s}$ are the fast and slow relaxation rates  governing the long-time behavior, respectively, and $A_{long}$ is asymmetry observed at longer times. 
The parameter $f$ measures the fractional contribution of the fast-relaxing component, while $(1 - f)$ corresponds to the fraction of the slow-relaxing component. \\ \hspace*{0.5 CM} Figure~\ref{Fig5}(d) shows the temperature dependence of $A_{\rm long}$, which follows a trend similar to that of the wTF asymmetry shown in Fig.~\ref{fig4}(b). $A_{\rm long}$ begins to decrease monotonically below $T_{\rm N1}$, indicating the emergence of a quasi-static magnetic order. Below $T_{\rm N2}$, it becomes nearly temperature-independent, suggesting the formation of a more stable static ordered state.
\\ \hspace*{0.5 CM}Moreover, the relaxation rate, $\lambda_{\mathrm{ZF}}^{f}$, shows a sharp increase as the system enters a magnetically ordered region below $T_{\rm N1} = 16$ K and begins decreasing as the temperature approaches $T_{\rm N2} = 11$ K, as shown in the inset of Fig.~\ref{Fig5} (d). A weaker, less conspicuous but similar behavior is observed for the $\lambda_{\mathrm{ZF}}^{s}(T)$ relxation rate. The divergent behavior of the relaxation rates near the magnetic transition temperatures is characteristic of the critical slowing down of spin fluctuations, which occurs as the system enters into an ordered magnetic state.  All these $\mu$SR results collectively suggest the occurrence of two successive magnetic transitions at $T_{\rm N1} = 16$ K and $T_{\rm N2} = 11$ K in $\mathrm{BiCrTeO}_6$, thus corroborating the bulk thermodynamic data. However, further neutron powder diffraction studies will be helpful in gaining deeper insight into its magnetic ground state and in elucidating the underlying microscopic spin structures in the AFM1 and AFM2 states.
 
\begin{figure}[h!]
\centering
\includegraphics[width = \columnwidth]{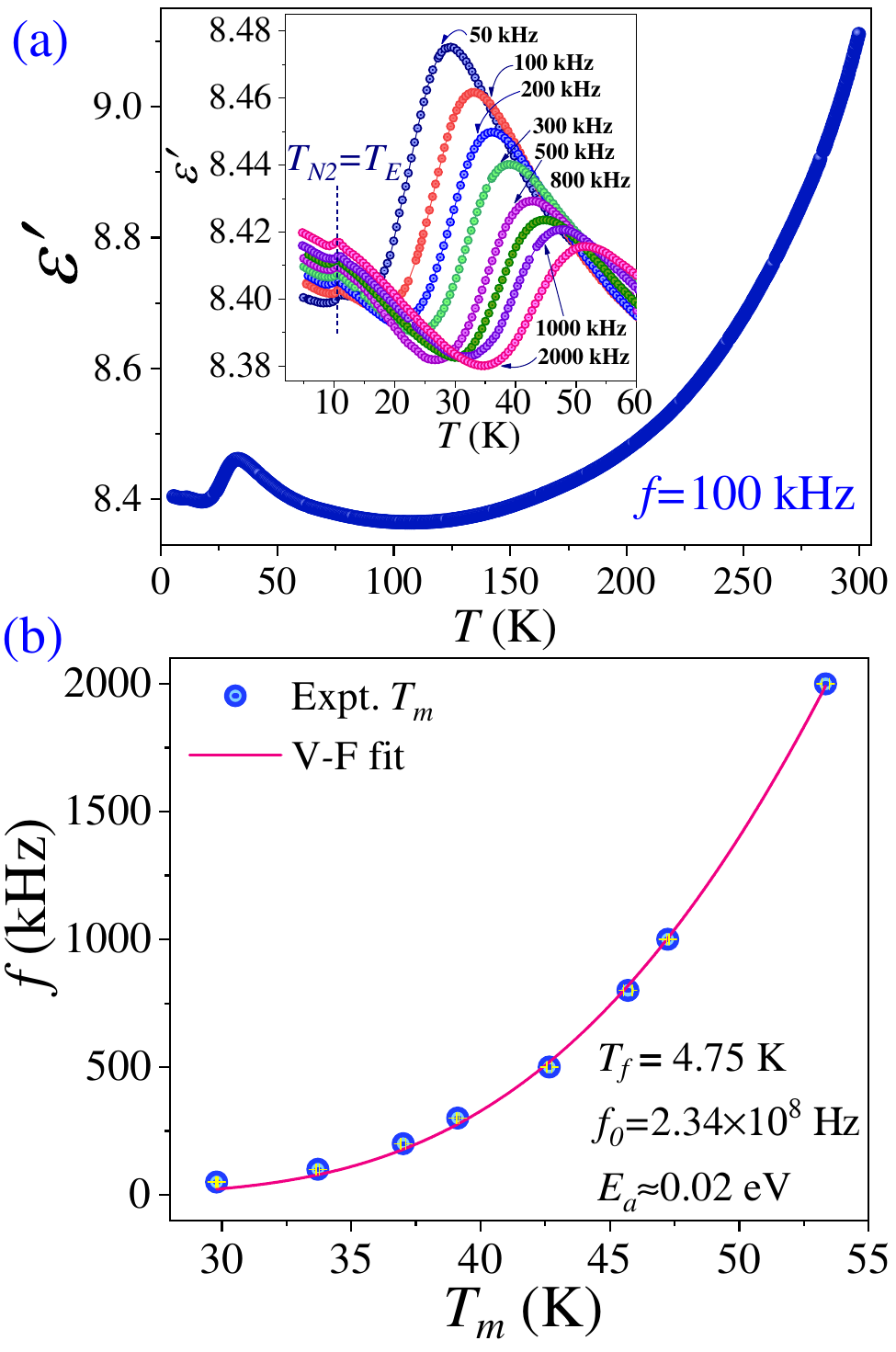}
\caption{(a) Thermal variation of the dielectric constant (\(\varepsilon'\)) of \(\mathrm{BiCrTeO}_6\) at a frequency \(f = 100\,\text{kHz}\). A closer view of the low-\(T\) region at various frequencies is shown in its inset. In the high-\(T\) region, \(\varepsilon'(T)\) exhibits a monotonic decrease with decreasing temperature, characteristic of lattice contraction. Upon cooling, a broad  \(f\)-dependent peak appears near \(T \approx 33\,\text{K}\), followed by a sharper  \(f\)-independent peak at \(T_{N2} = 11\,\text{K}\). (b) Vogel-Fulcher fit to the \(f\)-dependence of the dielectric maximum temperature (\(T_m\)) for the relaxor peak.}
\label{Fig6}
\end{figure}

\subsection{Magnetodielectric coupling and multiferroic properties}
The thermal variation of the dielectric constant (\(\varepsilon'\)) of BiCrTeO$_6$ for a frequency \(f = 100\,\text{kHz}\) is shown in Fig. \ref{Fig6}(a). In the high-\(T\) region, \(\varepsilon'(T)\) exhibits a monotonous decrease with lowering temperature, which is typical behavior of a system manifesting lattice contraction. As the temperature is decreased, a relatively broad peak is observed at \(T \approx 33\,\text{K}\), which is followed by a smaller and sharper peak at \(T_{N2} = 11\,\text{K}\). To further investigate the nature of these dielectric peaks, the \(f\)-dependence of the \(\varepsilon'(T)\) curves were studied in the low-\(T\) regime, as shown in the inset of Fig.\ref{Fig6}(a). Interestingly, the two peaks have distinctly different characters: the higher-temperature broad peak shows strong frequency dispersion, whereas the lower-temperature sharp peak is frequency-independent.

\begin{figure}[h!]
\centering
\includegraphics[width = \columnwidth]{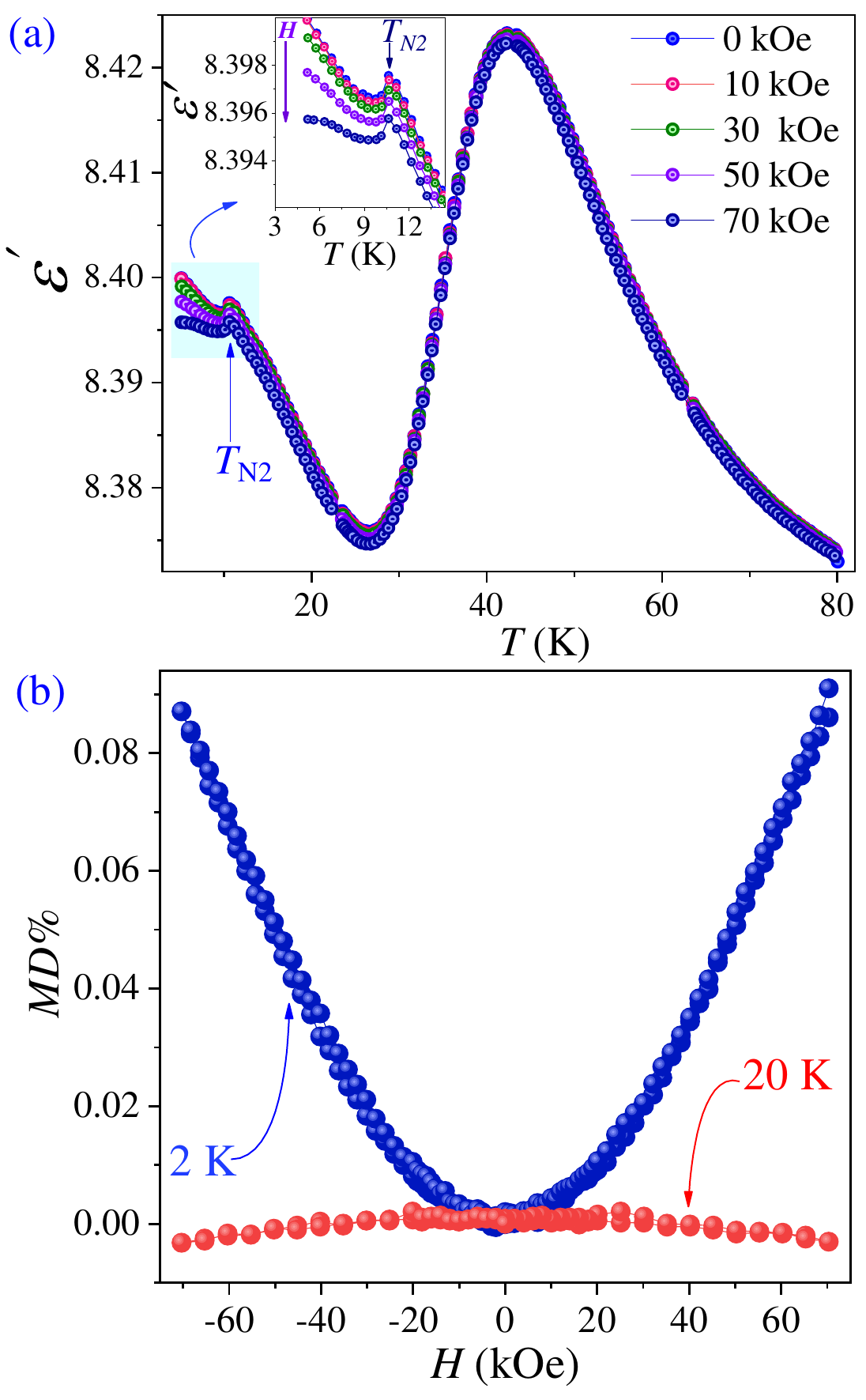}
\caption{(a) Thermal variation of the dielectric constant $\varepsilon'$ of $\mathrm{BiCrTeO}_6$, at a frequency $f = 100\,\text{kHz}$, in the low-$T$ region at various magnetic fields. The inset shows an expanded view of the region around $T_{N2} = 11\,\text{K}$. (b) Magnetodielectric effect as a function of applied magnetic field, measured at 2 K and 20 K.}

\label{Fig7}
\end{figure}

\begin{figure}[h!]
\centering
\includegraphics[width = \columnwidth]{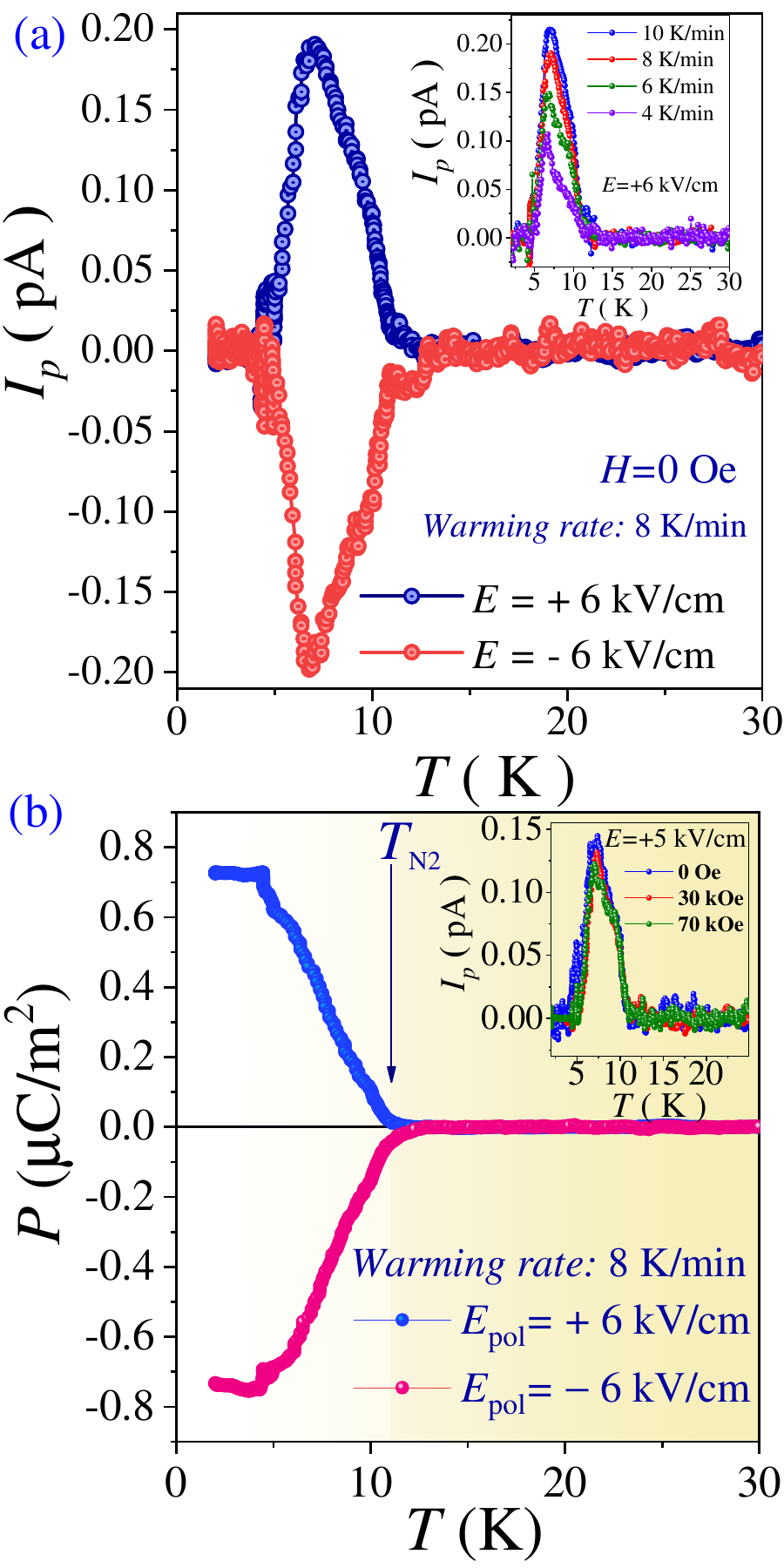}
\caption{(a) Pyrocurrent (\(I_p\)) curves under \(H\) = 0 Oe after poling the sample with a bias electric field of \(\pm 6\, \mathrm{kV/cm}\). Inset: \(I_p(T)\) curves collected for different warming rates under \(H\) = 0 Oe. (b) Ferroelectric polarization (\(P\)) as a function of temperature, corresponding to the \(I_p\) curves in (a). Inset: \(I_p(T)\) curves measured under various \(H\). }
\label{Fig8}
\end{figure}

The strong \(f\)-dependence of the higher-temperature peak position, along with its decreasing peak amplitude with increasing frequency, are typical features of a relaxor system \cite{relaxorg1,relaxorg2,relaxorg3}. As previously mentioned, \(\mathrm{BiCrTeO_6}\), exhibits substantial site disorder (approximately $~38\%$) involving two heterovalent ions, namely Cr\(^{3+}\) and Te\(^{6+}\), which serves as a probable source of relaxor behavior \cite{relaxorg2}. This relaxor behavior is triggered by the formation of clusters of polar nano regions (PNRs) due to quenched disorder introduced by anti-site disorder. This can be described by the random electric field model established for the prototypical relaxor material \(\mathrm{PbMg_{1/3}Nb_{2/3}O_3}\) and many related systems \cite{relaxorg2,relaxorg3}.

The \(f\)-dependence of the peak-maximum temperature (\(T_m\)) of \(\mathrm{BiCrTeO_6}\) was analyzed using a Vogel-Fulcher (VF) model, expressed as:
\[
f = f_0 \exp\left[\frac{E_a}{k_B (T_m - T_f)}\right]
\]
which is commonly followed by most relaxor materials \cite{vogel1921}. Here, \(E_a\) is the activation energy, \(f_0\) is the characteristic frequency, and \(T_f\) is the freezing temperature beyond which cluster reorientation ceases to be thermally activated. The best fit, as demonstrated in Fig. \ref{Fig6}(b), yields \(E_a \approx 0.02\, \text{eV}\), \(f_0 \approx 2.34 \times 10^8\, \text{Hz}\), and \(T_f \approx 4.75\, \text{K}\), values that are consistent with those reported for other relaxor systems \cite{relaxorg3}.
\\
\hspace*{0.5cm} In contrast, the \(f\)-independent and sharp, \(\lambda\)-like nature of the lower-temperature $\varepsilon'$ peak are typical characteristics of ferroelectric ordering \cite{multiferro1,multiferro2,BCBO}. Moreover, the corresponding dielectric loss (\(\tan \delta \sim 0.004-0.006\)) remains insignificant across the entire region of interest, ruling out any contribution from extrinsic charge carriers or spurious experimental effects \cite{MGTO,BCBO,MSO_PRB,multiferro1,FPO317_PRB}. The simultaneous occurrence of AFM spin ordering at \(T_{N2}\) clearly indicates a close correlation between these two phenomena. The $\varepsilon'(T)$ curves were recorded under various applied magnetic fields \(H\), as shown in Fig.~\ref{Fig7}(a). As \(H\) increases, a clear and systematic decrease in \(\varepsilon'\) is observed below \(T_{N2}\), thereby validating the presence of cross-coupling between the spin and dipolar order parameters. This magnetodielectric (MD) effect is more clearly visible in the inset of Fig.~\ref{Fig7}(a). To better understand the MD effect, isothermal $\varepsilon'$ versus \(H\)- curves were collected at \(T=2\,\mathrm{K}\) and \(20\,\mathrm{K}\). The corresponding $\text{MD\%} = \left[ \frac{\varepsilon'(0) - \varepsilon'(H)}{\varepsilon'(0)} \right] \times 100\%$ was subsequently calculated and plotted in Fig.~\ref{Fig7}(b), which demonstrates a pronounced MD effect at 2 K (\(< T_{N2}\)), while it is negligible at 20 K (\(> T_{N2}\)).
\\\hspace*{0.5cm}To verify whether the dielectric peak at \(T_{N2}\) is accompanied by ferroelectric polarization, pyrocurrent (\(I_p\)) measurements were performed. The sample was cooled through \(T_{N2}\) under a poling electric field of \(E_p = \pm 6\, \mathrm{kV/cm}\), followed by a 30-minute waiting period after short circuiting to nullify stray charges. Subsequently, the \(I_p(T)\) curves were measured during heating (under \(H = 0\, \mathrm{Oe}\)), as shown in Fig.~\ref{Fig8}(a). A clear \(I_p\) peak with an asymmetric shape appears just below \(T_{N2}\). This \(I_p\) signal is reversed, displaying a symmetrically opposite peak for the opposite poling direction, indicating a reversible ferroelectric polarization (\(P\)). The corresponding polarization versus temperature curve, shown in Fig.~\ref{Fig8}(b), indicates that ferroelectric order emerges below \(T_{N2} = 11\,\text{K}\).

\begin{figure}
\centering
\includegraphics[width = \columnwidth]{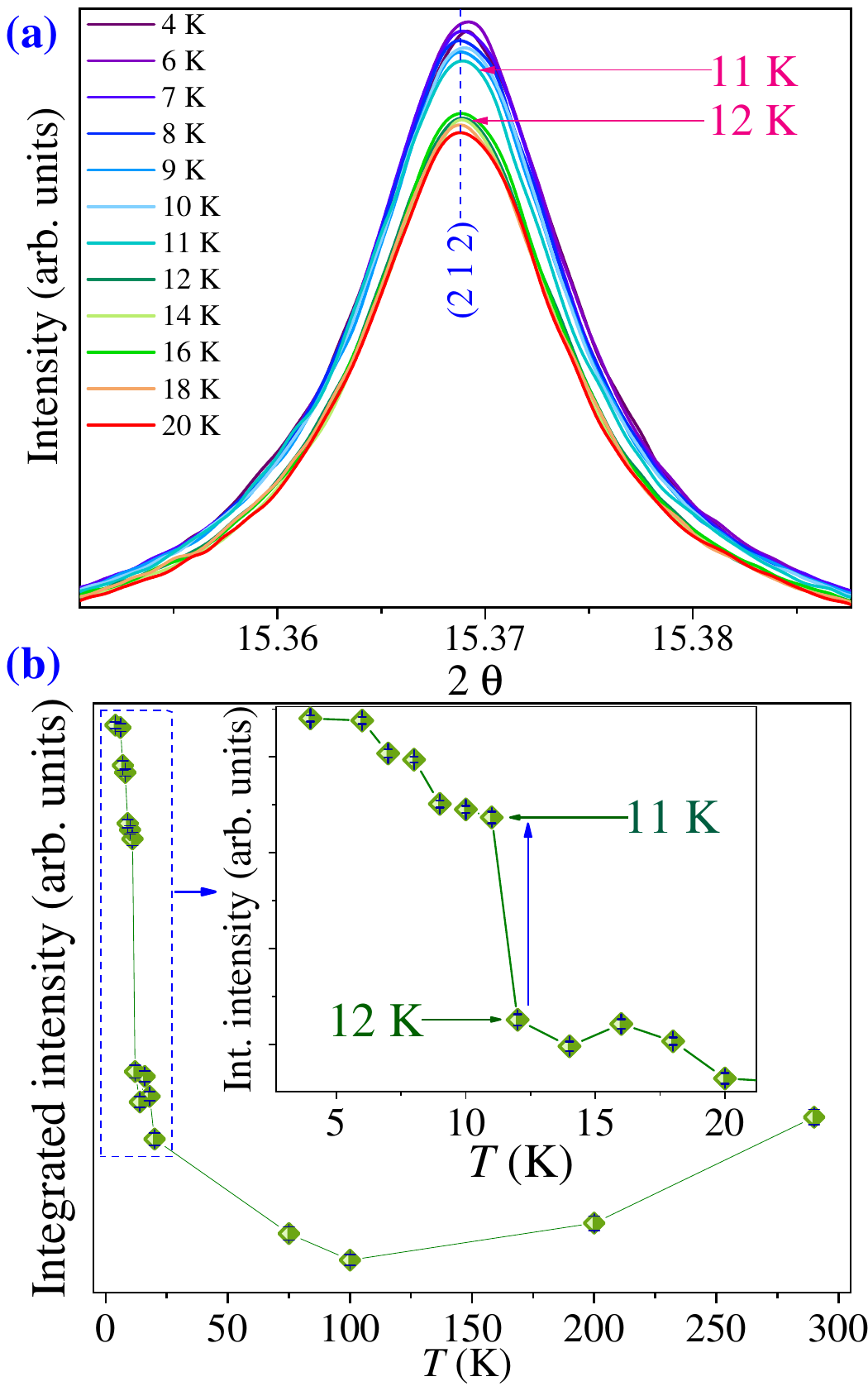}
\caption{(a) Selected diffraction peak $(2\ 1\ 2)$ measured at various temperatures shows an abrupt change in intensity at $T_{N2} = 11\,\text{K}$.
(b) Temperature variation of integrated intensity of $(2\ 1\ 2)$ peak.}
\label{Fig9}
\end{figure}

This suggests that the origin of the ferroelectric order is intertwined with the AFM spin ordering occurring below \(T_{N2} = 11\,\text{K}\), likely through magnetostructural coupling, as supported by our SXRD studies (discussed in the following section). The saturation value of \(P\) is approximately \(0.75\, \mu\mathrm{C/m^2}\). The \(I_p(T)\) curves were recorded at different warming rates, as plotted in the inset of Fig.~\ref{Fig8}(a). The pyrocurrent peak does not show any noticeable shift in position, thus confirming the intrinsic nature of the observed ferroelectric ordering \cite{multiferro1}. The \(I_p(T)\) measurements were also conducted under different magnetic fields \(H\), as shown in the inset of Fig.~\ref{Fig8}(b). A slight decrease in the \(I_p\) signal is observed as \(H\) increases, indicating weak magnetoelectric (ME) coupling. 
\\
\hspace*{0.5cm} The relatively small value of \(P\) in \(\mathrm{BiCrTeO_6}\) suggests that it is a weakly ferroelectric material. Similar small values of \(P\) have been reported in many other multiferroic compounds \cite{multiferro_smallP}. In fact, the \(I_p\) signal is often suppressed in polycrystalline samples compared to their single-crystal counterparts \cite{poly_single_pyro_ref}. Therefore, further studies on single crystals of \(\mathrm{BiCrTeO_6}\) may offer deeper insights into its ferroelectric properties.

\begin{figure}
\centering
\includegraphics[width = \columnwidth]{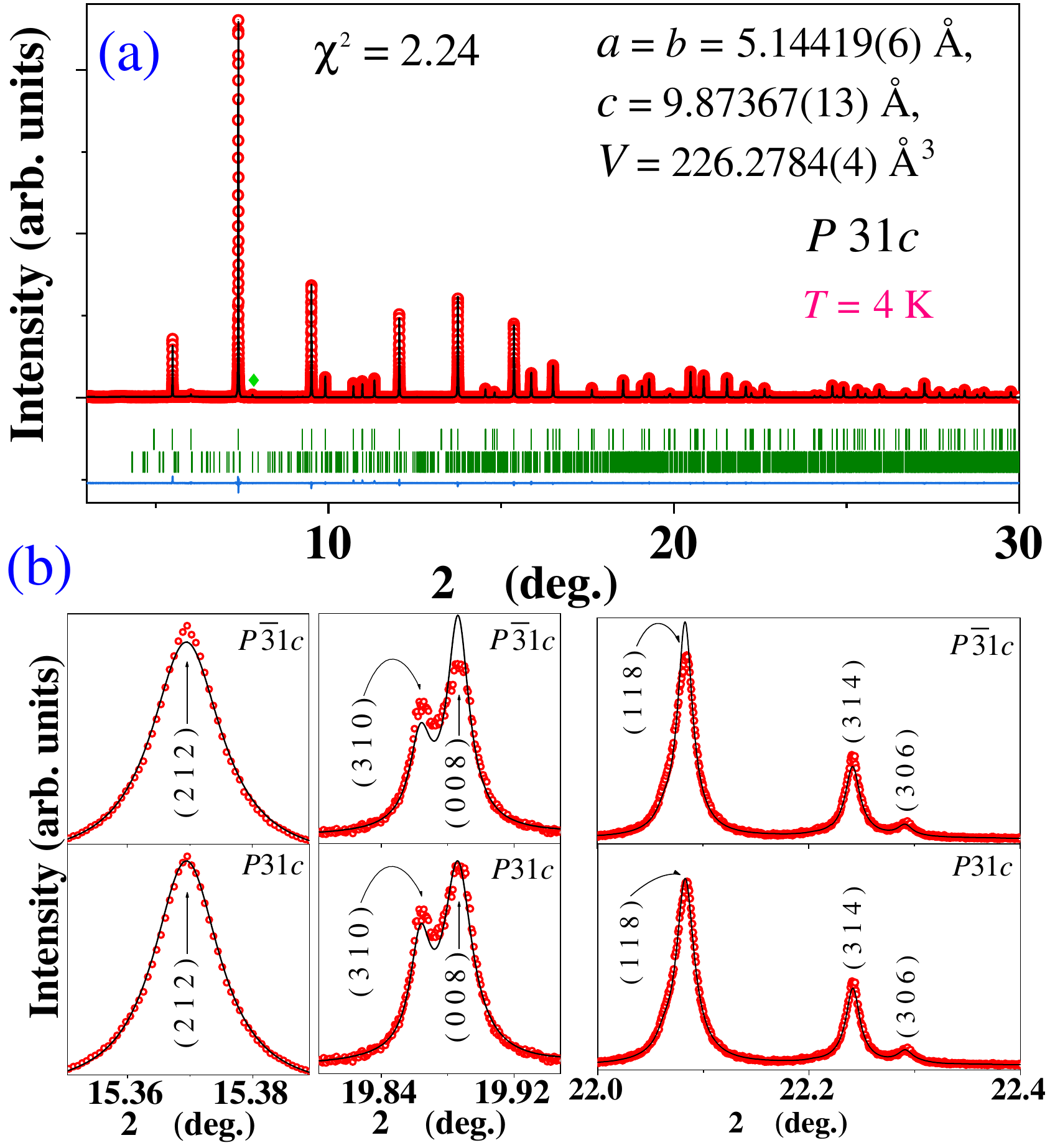}
\caption{
(a) SXRD pattern collected at $T = 4\,\text{K}$ (red data points) along with its fitted profile (black) using Rietveld refinement with the symmetry $P31c$.
The top and bottom green vertical bars indicate the allowed Bragg peak positions for the main phase $\mathrm{BiCrTeO_6}$ and the minor impurity phase $\mathrm{Bi}_{6}\mathrm{Te}_{2}\mathrm{O}_{15}$, respectively. A weak impurity phase peak is indicated by a green $\Diamond$ symbol.
(b) Closer views of selected Bragg peaks, along with their Rietveld-refined profiles using two different space groups, $P\bar{3}1c$ (upper panel) and $P3_1c$ (bottom panel). The fits are clearly better for $P31c$ symmetry.}
\label{Fig10}
\end{figure}

\subsection{Temperature-dependent synchrotron X-ray diffraction}
The crystal structure of a compound, is a key factor governing the dielectric and ferroelectric properties of multiferroic systems \cite{daniel_2009,SM_STO,multiferro2,BCBO,SGiri_CRO,MGTO,CTMO}. The structural evolution of \(\mathrm{BiCrTeO_6}\) at low temperatures, particularly near \(T_{N2} = 11\,\text{K}\), is crucial for understanding the origin of the observed ferroelectric order.

\begin{figure*}[htbp]
    \centering
    \includegraphics[width=\textwidth]{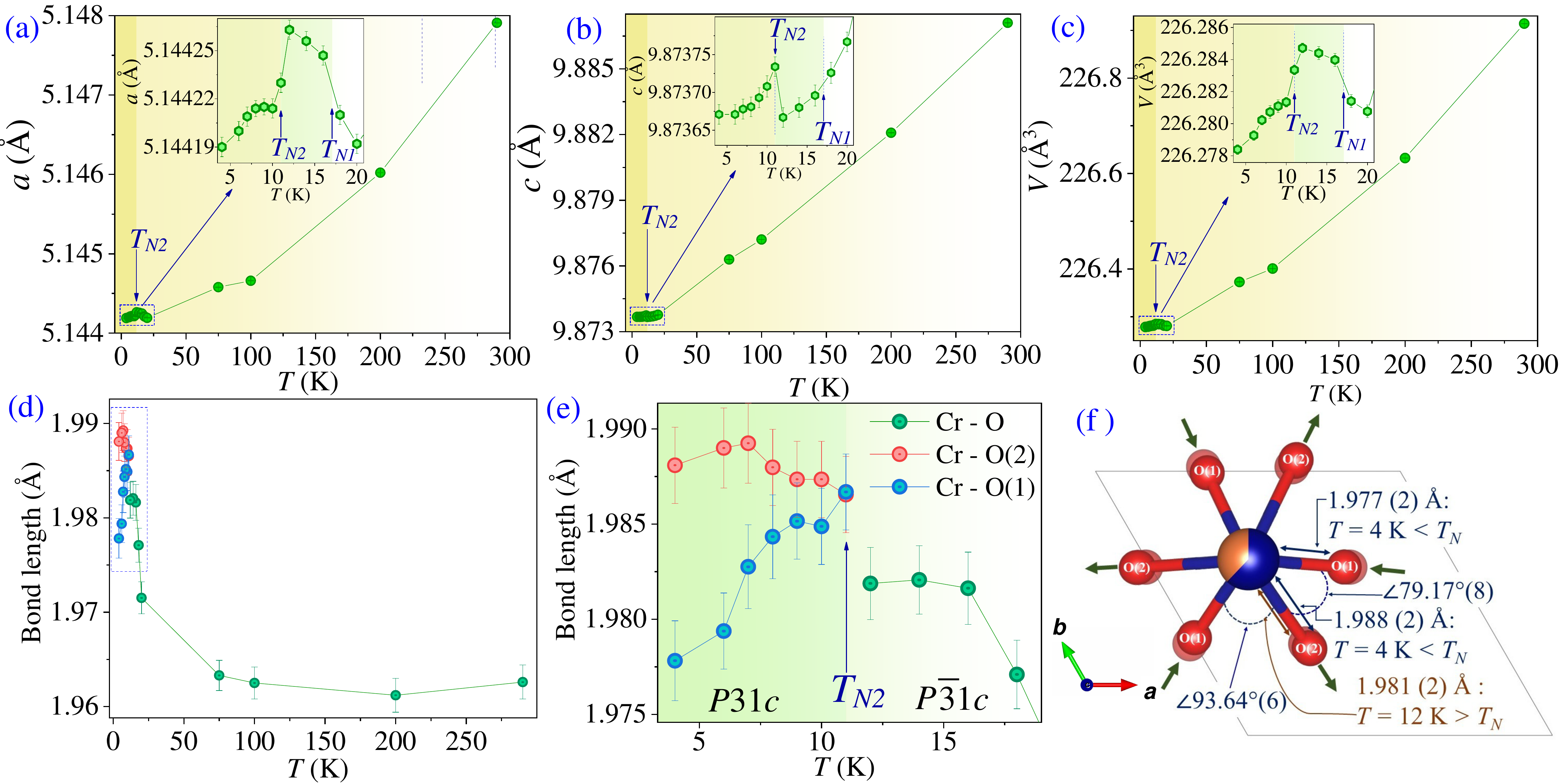}
    \caption{(a), (b), (c) Temperature dependence of the lattice parameters \textit{a = b}, \textit{c}, and the unit cell volume \textit{V}, respectively. The insets provide closer views of the corresponding curves near $T_{N2}$. (d) Variation of Cr–O bond lengths as a function of $T$, highlighting an octahedral distortion that occurs below $T_{N2}$. This is more clearly illustrated by the close-up view of the curve near $T_{N2}$ (e). (f) Corresponding structural distortion, involving the contraction of Cr–O(1) bonds and the elongation of Cr–O(2) bonds within the CrO$_{6}$ octahedra.}
    
    \label{Fig11}
\end{figure*}

The SXRD patterns measured at various temperatures are shown in Fig. S1 of the Supplementary Material. A closer examination reveals that several Bragg peaks display a significant and sudden change in intensity as the temperature crosses \(T_{N2} = 11\,\text{K}\). This behavior is illustrated in Fig.~\ref{Fig9}(a) for the \((2\, 1\, 2)\) peak within a narrow 2\(\theta\) range, and similar behavior is observed in several other peaks. The accompanying plot in Fig.~\ref{Fig9}(b) shows the temperature dependence of the integrated intensity of this peak, which shows an abrupt increase at $T_{N2}$. Our fitting of the SXRD patterns collected below \(T_{N2} = 11\,\text{K}\) using $P\bar{3}1c$ symmetry yielded unsatisfactory results, with various peaks at higher 2$\theta$ values displaying significant discrepancies in their observed and fitted intensities, as illustrated in the upper panels of Fig.~\ref{Fig10}(b). This possibly suggests that a structural phase transition might occur at $T_{N2}$.

\hspace*{0.5 CM} The ISODISTORT \cite{ISODISTORT} software was used to identify possible subgroups of the space group $P\bar{3}1c$ (163), among which $P31c$ (159) exhibits the highest symmetry. Subsequently, attempts to fit the data for $T \leq T_{N2}$ resulted in satisfactory fits. This is illustrated in Fig.~\ref{Fig10}(a) and the bottom panels of Fig.~\ref{Fig10}(b) and strongly suggests that a structural phase transition lowers the symmetry to the polar space group $P31c$ below $T_{N2}$. This is consistent with the emergence of ferroelectric order, likely due to a magnetostriction effect. Notably, a similar symmetry lowering from $P\bar{3}1c$ (163) to $P31c$ (159), which induces ferroelectricity, was previously reported for the well-known multiferroic lamellar system CuInP$_{2}$Se$_{6}$ \cite{CuInPSe1,CuInPSe2}.

 Detailed information on the refined crystallographic parameters of \(\mathrm{BiCrTeO_6}\) at 290 K and 4 K can be found in the Supplemental Materials \cite{Supplemental}.

  \hspace*{0.5 CM} The temperature dependence of the refined lattice parameters \textit{a = b}, \textit{c}, and the unit cell volume \textit{V} are shown in Figs.~\ref{Fig11} (a), (b), and (c), respectively. The usual lattice contraction with decreasing temperature is exhibited down to at least $T$ = 20 K. However, the curves \textit{a (T)} and \textit{V (T)} exhibit pronounced anomalies below $T_{N1}$, suggesting a subtle structural distortion that may be mediated by magnetostructural coupling. A monotonic decrease with $T$ across $T_{N1}$ is observed for the \textit{c (T)} curve. All three curves show clear anomalies near $T_{N2}$, which indicates a pronounced structural distortion. A slight increase in \textit{c (T)} below $T_{N2}$ suggests an increase in the inter-planar distance between the honeycomb layers. However, the decrease in volume suggests that a magnetostriction effect takes place. 
  \\ \hspace*{0.5CM}To gain deeper insight into the nature of this phase transition, the Cr-O bond lengths associated with the CrO$_{6}$ octahedra were studied as a function of temperature, as shown in Figs.~\ref{Fig11} (d) and ~\ref{Fig11}(e). The $P31c$ structure possesses two distinct Wyckoff positions for the oxygen atoms, O(1) and O(2). It is observed that the bond-length Cr-O(1) contracts below $T_{N2}$, whereas Cr-O(2) becomes elongated. This gives rise to more distorted CrO$_{6}$ octahedra, as schematically illustrated in Fig. ~\ref{Fig11} (f). This octahedral distortion may essentially trigger ferroelectric polarization within this polar structure. It is plausible that the structural phase transition alleviates spin frustration associated with the honeycomb layers.

 Similar scenarios, involving the release of spin frustration through structural phase transitions, have been reported in other spin-frustrated multiferroics \cite{release1,release2,jeku_1}. However, a microscopic spin structure study of this system using neutron diffraction could provide further insights into the origin of the spin-driven ferroelectricity observed in \(\mathrm{BiCrTeO_6}\). Nevertheless, the strong interplay among spins, lattice, and electric dipoles in the present spin-frustrated system \(\mathrm{BiCrTeO_6}\) is manifested in a wide range of intriguing phenomena, including spin-driven ferroelectricity, making it a highly promising system for further experimental studies on single crystals and for detailed theoretical investigation.

\section{Conclusion}

In conclusion, we have conducted a systematic study on the layered honeycomb lattice antiferromagnetic compound BiCrTeO$_{6}$, using a suite of advanced characterization techniques. Our comprehensive investigations reveal two successive magnetic transitions at $T_{\rm N1}$ = 16 K and $T_{\rm N2}$ = 11 K, below which the system develops long-range antiferromagnetic ordering, as confirmed by bulk magnetization and specific heat measurements. These findings are further corroborated by local-probe $\mu$SR measurements, which track the evolution of the internal magnetic fields below $T_{\rm N1}$ and provide evidence of magnetic ordering. The appearance of Bessel-function-like zero-field muon spin polarization further suggests a complex magnetic ground state in this system. Notably, the magnetic transition at $T_{\rm N2} = 11$ K is accompanied by the emergence of a ferroelectric order. Additionally, the compound demonstrates dielectric relaxor behavior at temperatures well above the magnetically ordered state, triggered by the intrinsic Cr/Te anti-site disorder inherent in the system. Driven by a strong magnetoelastic coupling, the system undergoes a structural phase transition at $T_{\rm N2} = 11$ K, characterized by a symmetry lowering from the non-polar space group \textit{P}$\overline{3}\,1\,c$ (163) to the polar space group \textit{P}$3\,1\,c$ (159), as evidenced by high-resolution SXRD studies. This structural change appears to play a pivotal role in inducing the observed ferroelectric order in BiCrTeO$_{6}$, positioning it as an intriguing new multiferroic candidate exhibiting spin-driven ferroelectricity.

\section{Acknowledgements}
 This work was supported by the European Commission through the Horizon Europe MSCA Postdoctoral Fellowship (Project No. 101110742) and by the National Science and Technology Council, Taiwan, under Grants Nos. NSTC 114–2112-M-110–019, NSTC 113–2112-M-110–006 and NSTC 113–2112-M-110–024. A.P. and G.B. gratefully acknowledge the European Synchrotron Radiation Facility (ESRF) in Grenoble, France, for granting access to their experimental facilities (proposal ID IH-HC-4009) used for high-resolution SXRD measurements. They also thank the Paul Scherrer Institute (PSI) in Switzerland for enabling their $\mu$SR experiments (proposal ID 20240778). Additionally, A.P. thanks Dr. S. Mukherjee from the Institute of Physics, Chinese Academy of Sciences, Beijing, China, for insightful discussions.

\bibliographystyle{PRB}
\bibliography{References.bib}
\end{document}